\documentclass[conference]{IEEEtran}
\IEEEoverridecommandlockouts
\usepackage{cite}
\usepackage{amsmath,amssymb,amsfonts}
\usepackage{algorithmic}
\usepackage{graphicx}
\usepackage{textcomp}
\usepackage{xcolor}
\usepackage{bm}
\usepackage{subfig} 
\def\BibTeX{{\rm B\kern-.05em{\sc i\kern-.025em b}\kern-.08em
		T\kern-.1667em\lower.7ex\hbox{E}\kern-.125emX}}
\begin{document}
	
\title{SC-CDM: Enhancing Quality of Image Semantic Communication with a Compact Diffusion Model}

\author{
	\IEEEauthorblockN{
		Kexin Zhang\IEEEauthorrefmark{1}, 
		Lixin Li\IEEEauthorrefmark{1}, 
		Wensheng Lin\IEEEauthorrefmark{1}, 
		Yuna Yan\IEEEauthorrefmark{1}, 
		Wenchi Cheng\IEEEauthorrefmark{2} 
		and Zhu Han\IEEEauthorrefmark{3}} 
	\IEEEauthorblockA{\IEEEauthorrefmark{1}School of Electronics and Information, Northwestern Polytechnical University, Xi’an, China, 710129}
	\IEEEauthorblockA{\IEEEauthorrefmark{2}State Key Laboratory of Integrated Services Networks, Xidian University, Xi’an, China, 710071}
	\IEEEauthorblockA{\IEEEauthorrefmark{3}Department of Electrical and Computer Engineering, University of Houston, Houston, TX, 77004} 
	\thanks{This paper has been accepted for publication in IEEE Globecom 2024 workshop.}
} 

\maketitle

\begin{abstract}
Semantic Communication (SC) is an emerging technology that has attracted much attention in the sixth-generation (6G) mobile communication systems. However, few literature has fully considered the perceptual quality of the reconstructed image. To solve this problem, we propose a generative SC for wireless image transmission (denoted as SC-CDM). This approach leverages compact diffusion models to improve the fidelity and semantic accuracy of the images reconstructed after transmission, ensuring that the essential content is preserved even in bandwidth-constrained environments. Specifically, we aim to redesign the swin Transformer as a new backbone for efficient semantic feature extraction and compression. Next, the receiver integrates the slim prior and image reconstruction networks. Compared to traditional Diffusion Models (DMs), it leverages DMs' robust distribution mapping capability to generate a compact condition vector, guiding image recovery, thus enhancing the perceptual details of the reconstructed images. Finally, a series of evaluation and ablation studies are conducted to validate the effectiveness and robustness of the proposed algorithm and further increase the Peak Signal-to-Noise Ratio (PSNR) by over 17\% on top of CNN-based DeepJSCC.
\end{abstract}


\begin{figure*}[!t]
	\centering
	\includegraphics[width=\textwidth]{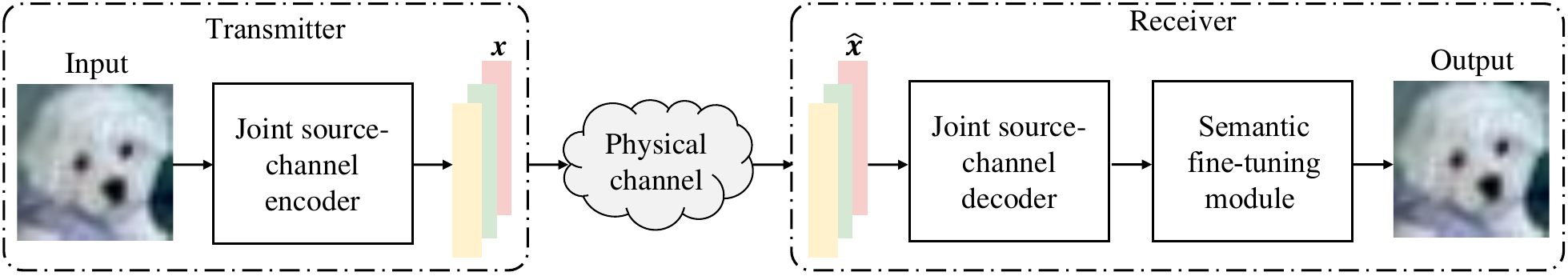}
	\caption{The overall architecture of the proposed SC-CDM system.}
	\label{fig1}
\end{figure*}

\section{Introduction}
With the rapid advancement of wireless communications, the connections between data and mobile devices are increasing exponentially. 
As 5G technologies approach the Shannon physical layer capacity limit, Semantic Communication (SC) is anticipated to be crucial for 6G networks \cite{ref1, Lin2024SIC, Lin2024SF}. 
Unlike traditional communication, SC integrates Deep Learning (DL) to extract core semantic information relevant to specific tasks directly from source data using Deep Neural Networks (DNNs) \cite{DL2021, ML2023}, which are then encapsulated in encoded features. 
Specifically, SC can effectively support diverse applications in 6G. 

In a seminal work \cite{ref2}, Deep Joint Source-Channel Coding (DeepJSCC) employs neural networks for its encoder and decoder, which are jointly trained to minimize Mean Square Error (MSE) and optimize Structural Similarity (SSIM). 
In related research \cite{ref3}, authors have enhanced image transmission by proposing transmitting high-level semantic information like text descriptions, particularly useful in bandwidth-limited scenarios. 
Furthermore, DeepJSCC has been adapted for digital communication systems as DeepJSCC-Q \cite{ref4}, improving its integration with existing communication infrastructures. JSCC approaches have shown outstanding performance due to their robust learning capabilities. However, these methods focus more on pixel-level distortion or structural similarity rather than perceptual distortion.

SC is also the fusion of communication and computing \cite{CDMR2022}, which has a high requirement for efficiency.
To address the challenges of scalability and efficiency in SC, 
Yu \emph{et al.} \cite{ref5} introduced a two-way semantic communication system that reduces training overhead by eliminating the need for information feedback and model migration. 
Nguyen \emph{et al.} \cite{ref6} developed a multi-user semantic communication system that adapts transmission lengths and employs hybrid loss optimization for high-quality image reconstruction and congestion avoidance. However, the absence of feedback and adaptation mechanisms complicates real-time adjustments, making practical deployment difficult.

Given the success of AI-Generated Content (AIGC), generative models can enhance transmission efficiency by creating complex multimedia content from low-dimensional data. A study \cite{ref6add} proposed the Latent Diffusion DNSC scheme, combining adversarial learning and variational autoencoders for intelligent online de-noising. Another technique \cite{ref7} implements semantic segmentation at the transmitter and uses Generative Adversarial Networks (GANs) for image reconstruction. However, GAN-based approaches \cite{ref8} often require significant computational resources and can face instability issues during training.

Considering the limitation of current SC schemes in failing to balance the effectiveness of generative models with the complexity of SC transmission, we design a generative semantic communication framework by exploiting the potential of generative models to bridge these gaps, aiming to enhance the transmission quality further. Now, we provide the main contributions of this paper, which are as follows:

\begin{itemize}
	\item To minimize the distortion difference between the reconstructed signal at the receiver and the source at the transmitter, we design a semantic communication architecture base on the swin Transformer. It avoids the need for precise bitstream recovery, greatly lowering communication costs.
	
	\item We employ a compact diffusion model within a generative semantic communication framework adept at reconstructing complex scenes from compressed semantic information. This model capitalizes on the advanced distribution mapping capabilities of Diffusion Models (DMs) to calculate a compact condition vector, enhancing efficient image recovery at the decoding stage. This approach significantly lowers computational demands relative to conventional diffusion models. 
	
	\item Employing the real datasets, simulation results show that our proposed framework significantly outperforms traditional separation-based schemes and further boosts PSNR by over 17\% on top of DeepJSCC. 
\end{itemize}

\section{System Model}

\subsection{Problem Formulation}

Communication must deal with physical challenges such as channel distortion and signal corruption or loss. First, given the limited transmit power of the transmitting device, the transmitted signal $\bm{x}$ must satisfy the average power constraint $\frac{1}{k}{\mathbb{E}}{_x}\left[ {\left\| x \right\|_2^2} \right] \le P$. Then, the signal is sent to the noisy channel with the transmission representation $\bm{\hat{x}} = \bm{A}\bm{x} + \bm{\varepsilon}$, where $\bm{\varepsilon} \sim {\cal N}\left( {0,{\sigma _n}^2} \right)$ is the noise added by the channel, and $\bm{A}$ is the matrix of the corruption, indicating the missing portions of the transmitted content. 
Therefore, the received content $\bm{\hat{x}}$ is a noisy and masked version of the original transmitted content $\bm{x}$. A quantitative method to characterize the amount of noise due to channel conditions is the PSNR.

\subsection{Semantic Communication Framework}
In this paper, we have designed a generative semantic communication network architecture, which includes a transmitter and a receiver. The main goal of the system is to reduce the number of bits transmitted to the user by exploiting the semantic information in the data while ensuring that the user maintains satisfactory performance despite the shortened length of the signal. 
Specifically, a swin Transformer-based codec rather than a conventional Vision Transformer (ViT) is first used as the backbone network for end-to-end semantic image transmission. This choice is based on several key enhancements of swin Transformer over ViT. Firstly, the computation complexity of swin Transformer scales linearly with image size thanks to its localised self-attention mechanism, as opposed to the quadratic scaling in ViT. Additionally, swin Transformer supports hierarchical representations through its patch merging technique, which allows for more nuanced feature processing. In our semantic encoder, we also adapted the patch sizes and dimensions post-merging to suit our needs better. 
Furthermore, we have integrated a target embedding vector directly into the self-attention blocks (the window-based multi-head self-attention and the shifted windows multi-head self-attention) to enhance the specificity and accuracy of the model in capturing relevant features. On the receiver side, the advantages of diffusion models in synthesising multimedia content are exploited to reconstruct high-quality images by preserving the transmitted semantic information. This method improves data transmission efficiency and optimises the ability to recover complex semantic content from lossy signals, thus significantly enhancing the overall system performance.

\begin{figure*}[!t]
	\centering
	\includegraphics[width=\textwidth]{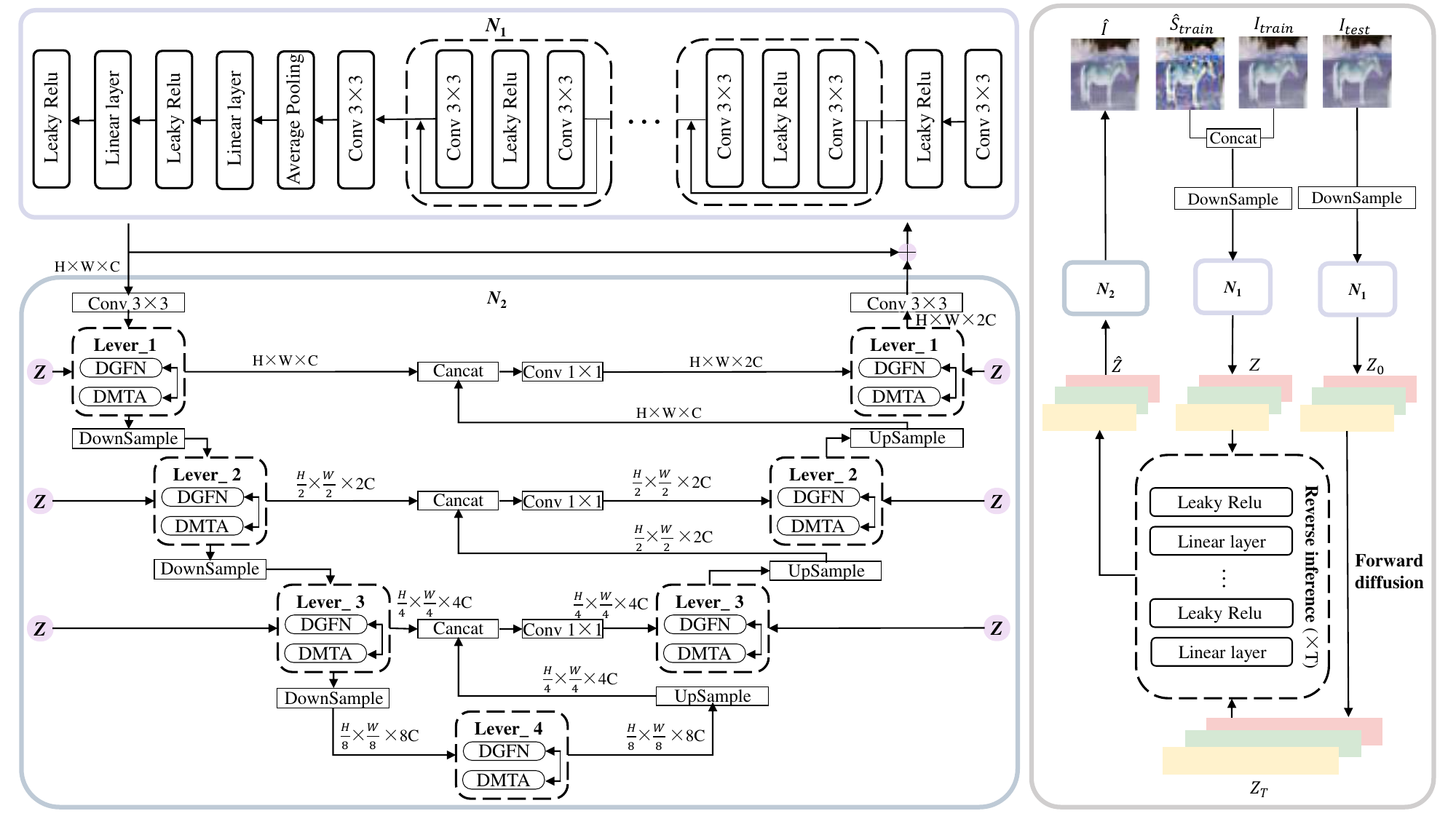}
	\caption{The pipeline of the semantic fine-tuning module pre-training, which consists of ${\bm{N}_1}$ and ${\bm{N}_2}$.}
	\label{fig2}
\end{figure*}

\section{Approach Details}

This paper proposes a generative semantic communication framework considering downlink transmission scenarios. Fig. \ref{fig1} shows the general framework of the SC-CDM. Specifically, the framework involves three main stages: the joint source-channel encoder, the joint source-channel decoder and the semantic fine-tuning module, described in detail in this section.

\subsection{Transmitter}

On the transmitter side, given an image set $I$, the images are first subjected to information mining and feature extraction. The captured key semantic information is encoded into the latent space denoted as $\bm{x} \in{\mathbb{R}} {^{\frac{H}{8} \times \frac{W}{8} \times C}}$ by the state-of-the-art vision transformer-swin Transformer \cite{ref9}. $H$ and $W$ indicate the height and width of the image, respectively. $C$ denotes the projected dimensions of the feature vector block.

During the training phase, considering the resolution of the test dataset, the decoder comprises two stages of transformer architecture to ensure that the model can capture rich visual features. 
Specifically, the first stage uses a $4\times4$ patch size to segment the image into multiple non-overlapping patches via the patch partition module and sequentially arranges these patches via linear embedding to form feature vector blocks. 
Then, these feature blocks are fed into the swin Transformer block. Swin Transformer block is a sequence-to-sequence function consisting of multiple swin Transformer layers. Each layer consists of a Window Multihead Attention layer and a Moving Window Multihead Attention layer, which are sequentially connected since the two sub-layers have the same input and output dimensions. Swin Transformer uses a shift-window division method within each layer to enhance modelling capabilities. It alternates the position of each window between successive layers, a strategy that further enhances the spatial perception of the model, improves signal immunity, and avoids the problem of window spacing boundaries. The second level focuses on detail capturing by splicing each set of 2×2 neighbouring patches through the patch merge layer, reducing the number of patch tokens to 1/4 of the original. Meanwhile, the dimension of patch tokens is expanded by a factor of four.
To reduce the output dimension, the downsampling of the feature map is achieved through a fully connected layer. 

After this process, the dimension of the spliced feature patch is reduced. This step-by-step processing from coarse to fine not only quickly identifies the main regions of the image but also digs deep into the details to improve the accuracy of feature extraction. In addition, the model uses the ReLU activation function between the input layer and the two hidden layers; the hyperparameter $C$ is set to 32, the loss function adopts the MSE, and the batch size is set to 32. the learning rate is set to 1×\(10^{ - 4}\), and the Adam optimiser is used for training.
It is worth noting that the semantic decoder follows an inverse architecture that is symmetric with the encoder.

\subsection{Receiver}

Inspired by \cite{ref10}, we develop the semantic fine-tuning module on top of an image restoration latent diffusion model, the core of which is a compact diffusion model. This model comprises two main networks: the slim prior network (${\bm{N}_1}$) and the image reconstruction network (${\bm{N}_2}$). ${\bm{N}_1}$ mainly extracts Prior Representations (PRs) to minimize the computational burden typical of traditional diffusion models. The image reconstruction network leverages semantic information from a pre-trained knowledge base to guide the production of high-quality images.

Fig. \ref{fig2} illustrates the semantic fine-tuning module pre-training workflow: during the \(\bm{N}_1\) training process, we initially concat the original training set images $\bm{\textit{I}}_{\textit{train}}$ with their corresponding decoded distortion images $\hat{S}_{\text{train}}$. 
Then, we downsample the merged images to serve as input for \(\bm{N}_1\). 
The extracted PRs from this process are designated as $Z$. 
Next, \(\bm{N}_2\) can use the extracted $Z$ to restore images, which are stacked with dynamic transformer blocks in the Unet shape. 
The dynamic transformer blocks consist of Dynamic Multi-head Transposed Attention (DMTA) and Dynamic Gated Feed-Forward Network (DGFN), which can use $Z$ as dynamic modulation parameters to add restoration details into feature maps, effectively aggregating both local and global spatial characteristics. 
Through joint optimization, the loss function is defined as \({{\cal L}_{1}} = \left\| \bm{\textit{I}}-\hat{\bm{\textit{I}}} \right\|_1\), where \(\bm{\textit{I}}\) and \(\hat{\bm{\textit{I}}}\) are the target and reconstructed images, respectively. \(\left\| \cdot \right\|_1\) denotes the ${\cal L}_{1}$ norm.

${\bm{N}_2}$ consists of two key sections: forward diffusion and reverse inference. Firstly, the PRs of the decoded image $\bm{\textit{I}}_{\textit{test}}$ captured by ${\bm{N}_1}$ denoted as $Z_0$, are used to apply the forward diffusion process of $Z_0$ to the sample $Z_T$ through $T$ iterations. Each iteration is as follows:

\begin{equation}
	\label{deqn_ex1a}
	q\left( Z_T \mid Z_0 \right) = \mathcal{N}\left( Z_T; \sqrt{\bar{\alpha}_T} Z_0, \left(1 - \bar{\alpha}_T\right)I \right),
\end{equation}

\noindent where ${\bar \alpha _t} = \prod\nolimits_{i = 1}^t {{\alpha _i}}$ is the cumulative product of ${\alpha _t}$, the scheduler gradually adds Gaussian noise at each time step $t \in \left[ {0,T} \right]$ until the semantic information of ${{Z_0}}$ becomes pure noise ${{Z_0}}$:

\begin{equation}
	\label{deqn_ex2a}
	{Z_t} = \sqrt {1 - {\beta _t}} {Z_{t - 1}} + \sqrt {{\beta _t}} \varepsilon,
\end{equation}

\noindent where $0 < {\beta _1} < {\beta _2} <  \cdots  < {\beta _T} < 1$ is a variance table with time-dependent constants and $\varepsilon$ is a Gaussian noise.

The forward diffusion process notifies the data; on the contrary, the backward inference is the denoising process. To take full advantage of the capability of ${\bm{N}_2}$, we perform all denoising iterations from a specific time step to obtain ${\hat Z}$ and send it to ${\bm{N}_2}$, thus enabling joint optimization with the denoising network ${\varepsilon _\theta }$. Then, the noise at each time step $t$ is estimated using ${\varepsilon _\theta }$, and ${\hat Z_{t - 1}}$ is obtained by:

\begin{equation}
	\label{deqn_ex3a}
	{\hat Z_{t - 1}} = \frac{1}{{\sqrt {{\alpha _t}} }}\left( {{{\hat Z}_t} - \frac{{1 - {\alpha _t}}}{{\sqrt {1 - {{\bar \alpha }_t}} }}{\varepsilon _\theta }\left( {{{\hat Z}_t},t} \right)} \right),
\end{equation}

\noindent where ${\varepsilon _\theta }$ is the prediction noise, ${\alpha _t} = 1 - {\beta _t}$, and ${\bar \sigma _t}$ is also a time-dependent constant for the step.

After $T$ iterations, \( \hat{Z} \in \mathbb{R}^{4C'} \) is generated. Since the pre-trained model only adds details for recovery, the compact diffusion model can obtain stable visual results after several iterations. After that, ${\bm{N}_2}$ utilizes \( \hat{Z} \) to recover the decoded image \( \hat{S} \).

The diffusion model aims to predict the noise distribution of the conditioned vectors of the decoded image with a loss function denoted as ${{\cal L}_{2}} = \frac{1}{4C'} \sum_{i = 1}^{4C'} \left| \hat{Z}(i) - Z_0(i) \right|$. To achieve efficient generation and propagation of semantic information, ${\cal L} = {{\cal L}_{1}} + {{\cal L}_{2}}$ is used here for joint optimization.

\section{Experiment Results}
\subsection{Experimental Setup}\label{AA}
We compare our proposal with classical separation-based source and semantic communication. The traditional communication model considers the well-established image codec JPEG. The channel noise or fading is then processed using a Low-Density ParityCheck code (LDPC) and Quadrature Amplitude Modulation (QAM) scheme, denoted as ”JPEG+LDPC+QAM”. The modulation order is set to 4. The semantic communication model employs the classical algorithm DeepJSCC. All methods consistently use the CIFAR-10 dataset and are selected to be trained with SNRs between 1 dB and 13 dB. Then, the tests are performed under Additive White Gaussian Noise (AWGN) and Rayleigh channel conditions, both under different channel conditions with PSNR values in the set [0, 3, 6, 9, 12, 15].

We adopt a four-level encoder-decoder structure during the compact diffusion model training process. From level 1 to level 4, the attention heads in DMTA are set to [1, 2, 4, 8], the number of dynamic transformer blocks to [3, 5, 6, 6], and the channel number $C'$ of ${\bm{N}_1}$ is set to 96. Training begins with a patch size of 32×32 and a batch size 16. In the second step, \(\beta _t\) linearly increases from 0.10 to 0.99, starting with an initial learning rate of 2×\(10^{ - 4}\), the total timesteps $T$ are set at 4, and the epoch is 30. In this simulation, the experimental platform for training and testing is built on an Ubuntu 20.04 system with CUDA 11.8 support, and the deep learning framework is Pytorch 2.0.0.

\begin{figure*}[!t]
	\centering
	\subfloat[]{\includegraphics[width=3.1in]{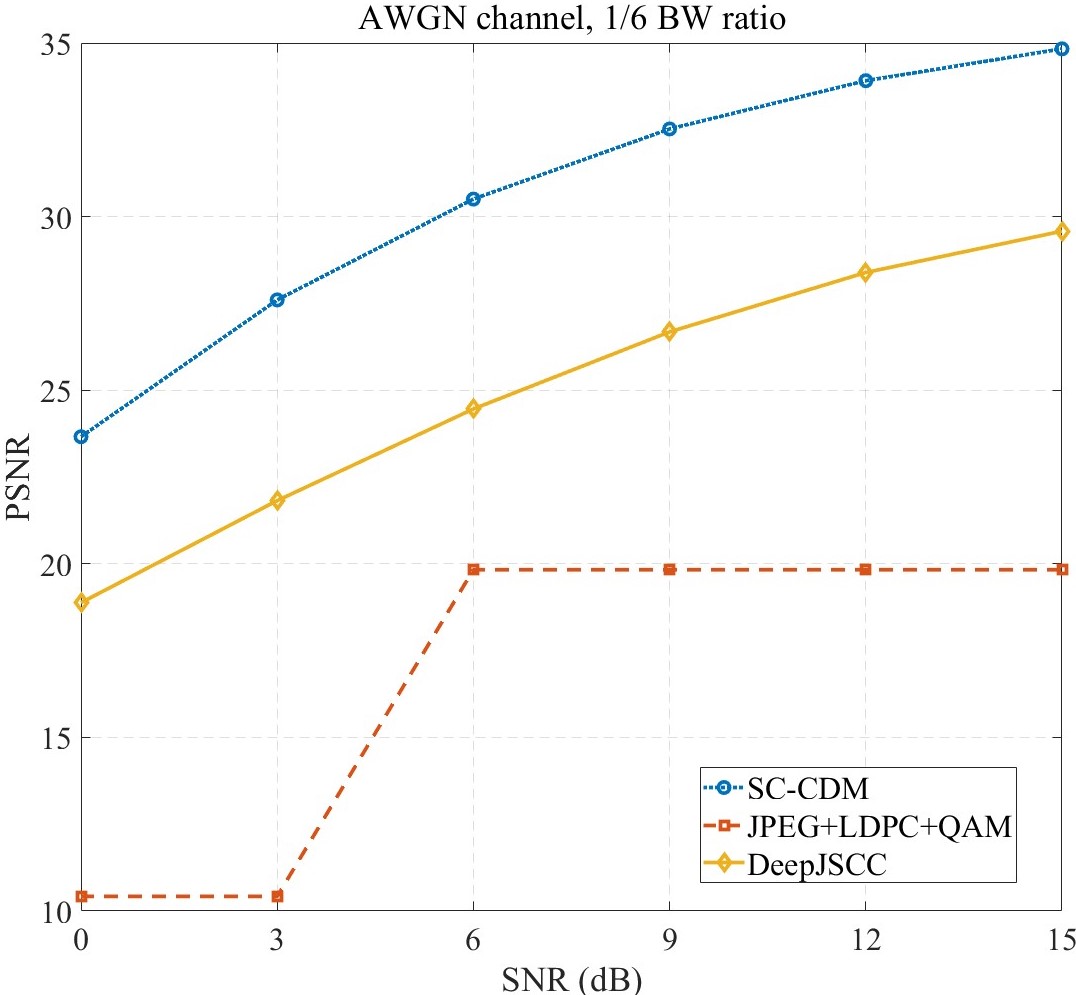}%
		\label{fig_first_case}}
	\hfil
	\subfloat[]{\includegraphics[width=3.1in]{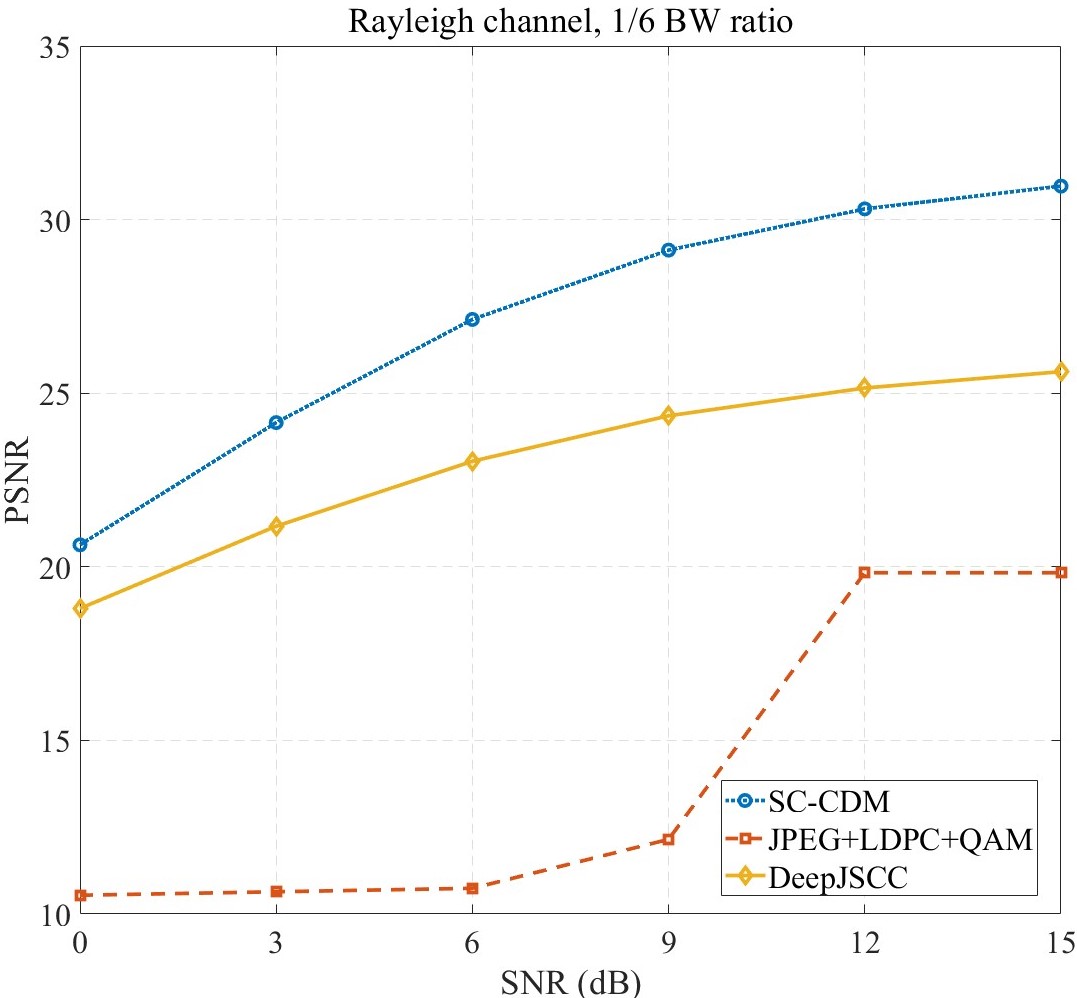}%
		\label{fig_second_case}}
	\caption{PSNR performance versus the SNR. (a) AWGN channel. (b) Rayleigh channel.}
	\label{fig3}
\end{figure*}

\subsection{Result Analysis}

\begin{figure*}[!t]
	\centering
	\subfloat[]{\includegraphics[width=3.1in]{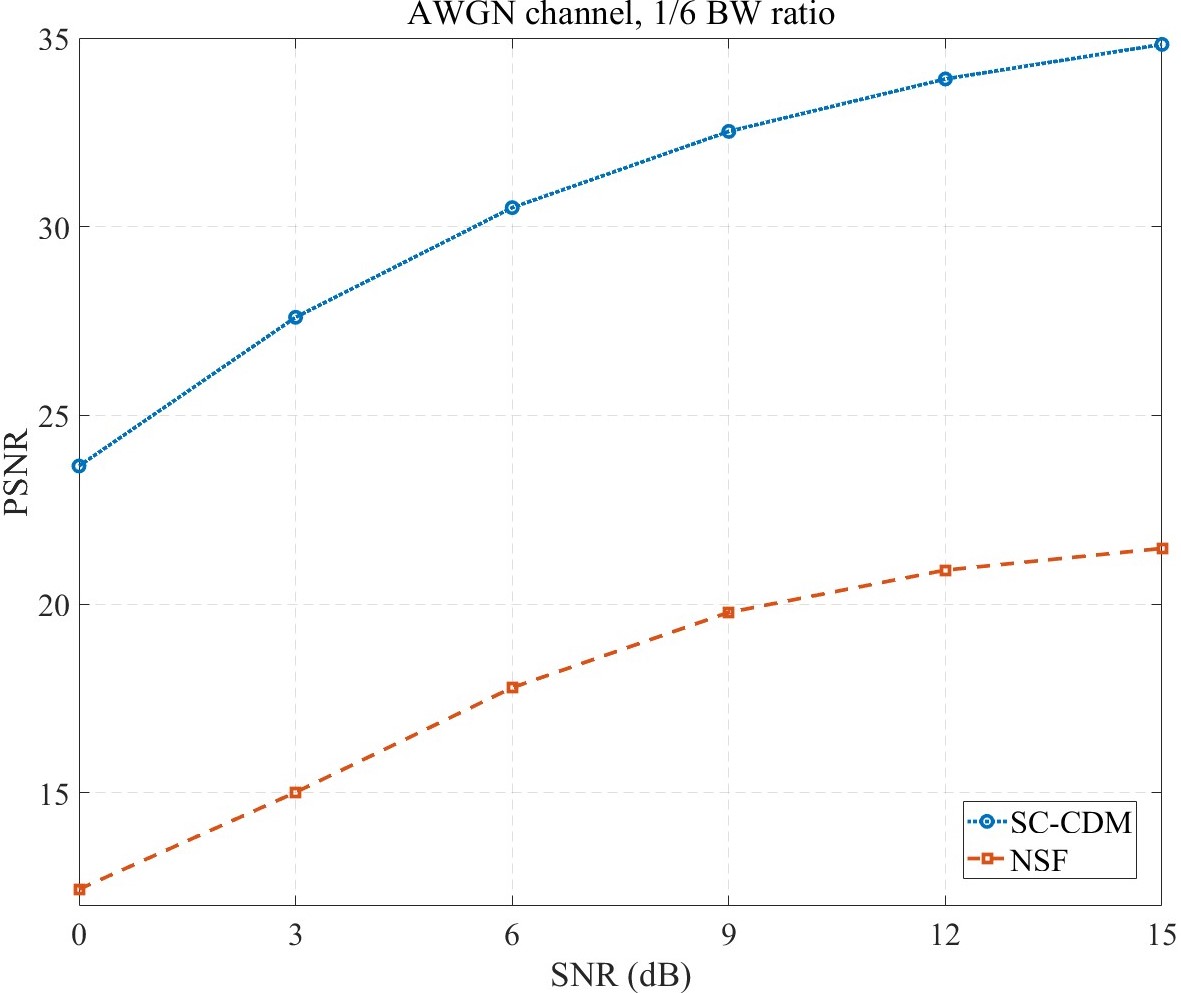}%
		\label{first_case}}
	\hfil
	\subfloat[]{\includegraphics[width=3.1in]{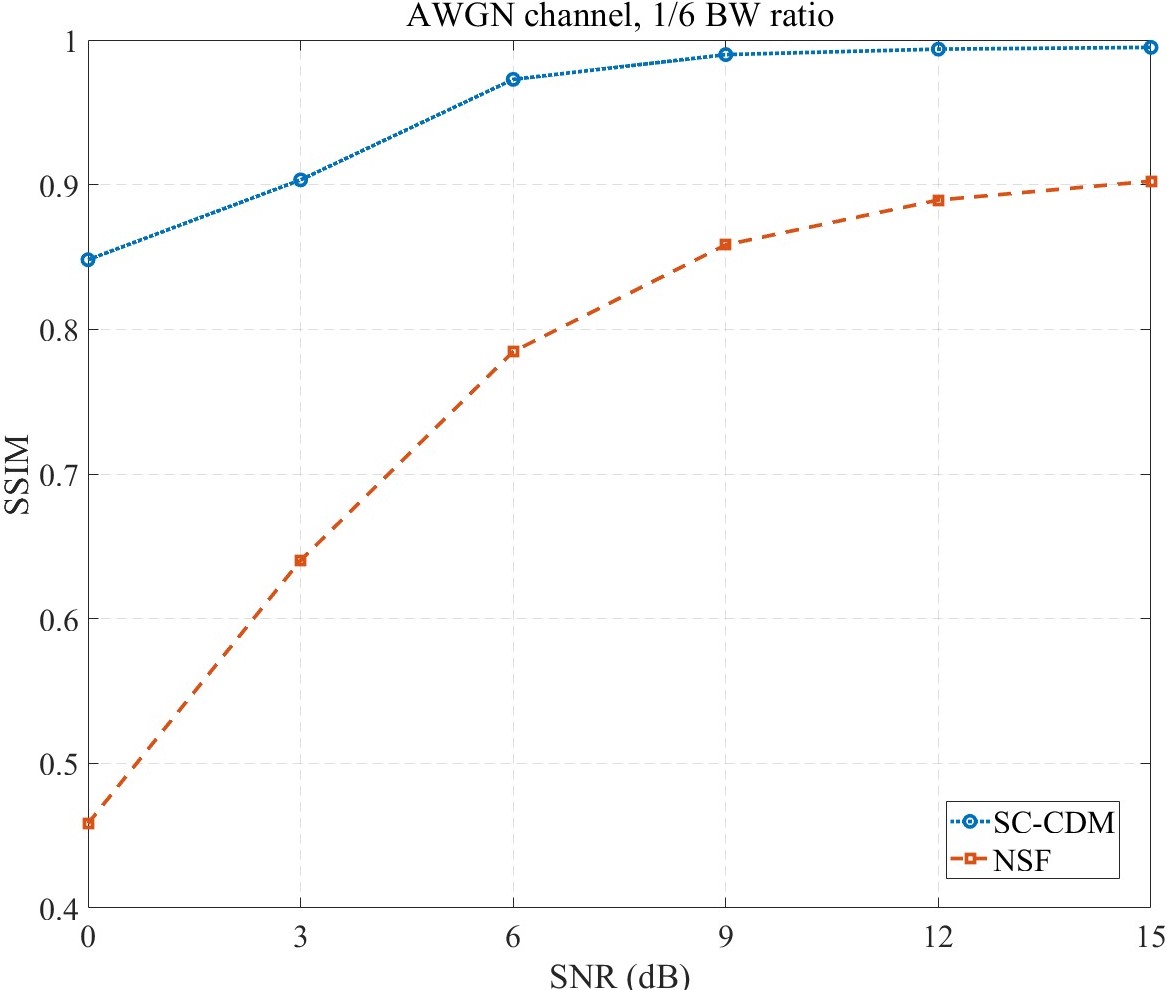}%
		\label{second_case}}
	\caption{The ablation performance versus the SNR. (a) PSNR. (b) SSIM.}
	\label{fig4}
\end{figure*}

\begin{figure*}[!t]
	\centering
	\subfloat[]{\includegraphics[width=1.75in]{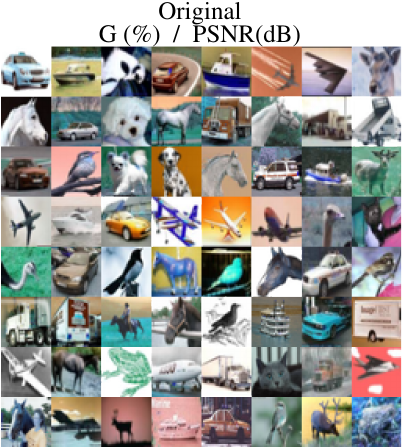}%
		\label{first_a}}
	\vspace{1ex}
	\subfloat[]{\includegraphics[width=1.75in]{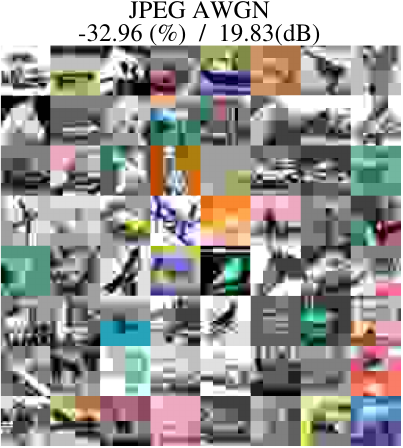}%
		\label{second_a}}
	\vspace{1ex}
	\subfloat[]{\includegraphics[width=1.75in]{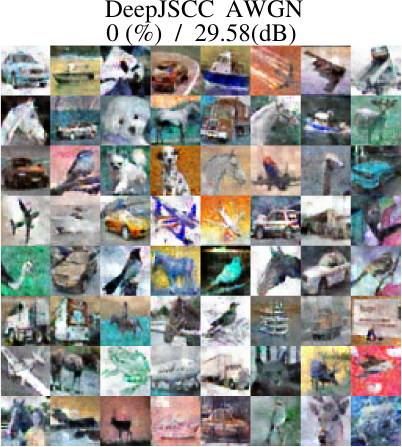}%
		\label{third_a}}
	\vspace{1ex}
	\subfloat[]{\includegraphics[width=1.75in]{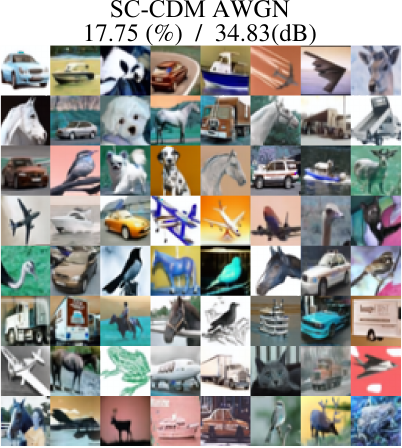}%
		\label{fourth_a}}
	\vspace{-1ex} 
	\subfloat[]{\includegraphics[width=1.75in]{F64_original.pdf}%
		\label{first_b}}
	\vspace{1ex}
	\subfloat[]{\includegraphics[width=1.75in]{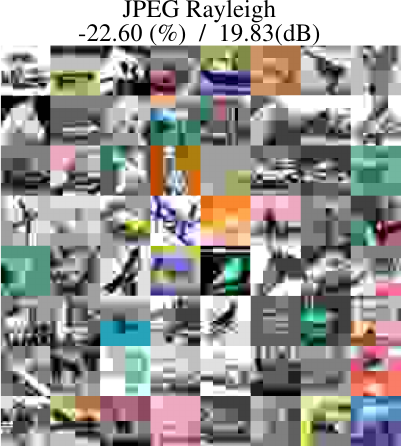}%
		\label{second_b}}
	\vspace{1ex}
	\subfloat[]{\includegraphics[width=1.75in]{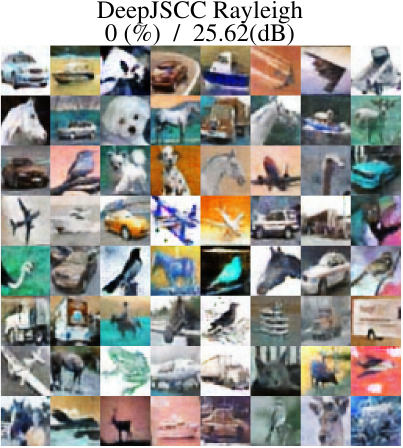}%
		\label{third_b}}
	\vspace{1ex}
	\subfloat[]{\includegraphics[width=1.75in]{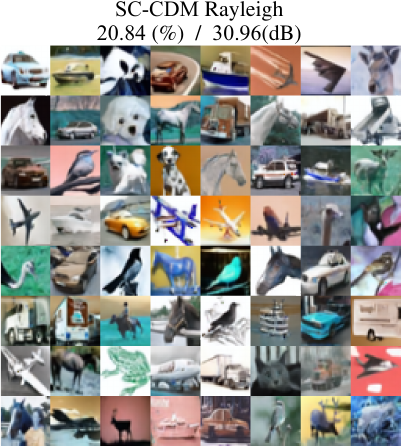}%
		\label{fourth_b}}
	\vspace{-2ex} 
	\caption{Visual comparison examples under two types of channels at SNR=15dB.}
	\label{fig5}
\end{figure*}

To verify the effectiveness, this section analyses the quality of images generated by three image communication methods under two different channel conditions. 

As reported in Fig. \ref{fig3}, we present the PSNR of the decoded image obtained using these approaches. For the SC-CDM, a single model covers a range of SNR from 0 dB to 15 dB. It consistently aligns with the expected trend of increased PSNR across both AWGN and Rayleigh channels as the SNR improves, and this technique demonstrates superior performance in all tested regions, highlighting its effectiveness in reducing redundancy and enhancing output quality. This advantage may be due to the limitations of JPEG compression, which, despite its prevalence, is lossy and can lead to irreversible information loss. In contrast, while LDPC codes and QAM modulation increase transmission robustness, they do not fully compensate for the quality degradation in the presence of errors.

In addition, in the forward channel (SNR ranges from 0 dB to 3 dB), compared to the transmission performance of JPEG, although the SC-CDM performance decreases as the SNR decreases, it does not degrade rapidly, demonstrating a significant graceful degradation advantage. Compared with DeepJSCC, the SC-CDM has a lower performance gap at lower SNR, indicating that even if the received semantic information image is severely corrupted, it can still generate a realistic image consistent with the original transmitted semantic information. Meanwhile, the PSNR of JPEG remains constant when the channel conditions are very bad or substantially improved, which reflects the so-called cliff effect. When the channel deteriorates beyond a threshold, the receiver cannot decode the channel code and, therefore, cannot transmit any semantic information. Comparatively, when the SNR is high, the PSNR reaches the performance saturation of conventional communication algorithms, and further enhancement will not improve the output quality.

To verify the validity of our methodological choices, we conducted ablation tests. Taking the AWGN channel as an example, Fig. \ref{fig4} shows the difference in performance when the semantic fine-tuning module is used (SC-CDM) during training and when this module is not used (NSF). The experimental results show that the SC-CDM can significantly improve the performance, confirming the excellent contribution of the compact diffusion model. These findings reinforce the potential of our approach for applications in image communication, especially in bandwidth-constrained scenarios where high-quality image reconstruction is required.


To provide a subjective quality assessment consistent with the human vision system. Taking SNR=15 as an example, Fig. \ref{fig5} provides a visualization example of the transmitted images in the three algorithms, and the percentage indicates the PSNR gain based on DeepJSCC. The results show that compared to the first column, the second column mainly demonstrates pixel-blocky reconstructed images; in the third and fourth columns, although there are errors in transmitting specific pixels, showing better visual effects. The proposed SC-CDM can acquire higher definition and accuracy image information by introducing the semantic fine-tuning module.

\section{Conclusion}
In this paper, we have proposed a high-efficiency scheme that focuses on transmitting extremely compressed content while ensuring high-quality semantic recovery at the receiver. Specifically, the system reduces the consumption of communication resources by building upon the swin Transformer and deploys a compact diffusion model in the decoder, which exploits the powerful multimedia generation capability of the diffusion model to improve the perceptual quality of the delivered images. Experiments show that the framework remains robust in reconstructing semantic information.

\vspace{12pt}
\color{red}
\end{document}